\documentclass[twocolumn,prb,superscriptaddress,preprintnumbers,amsmath,amssymb]{revtex4}
\usepackage[dvips]{graphicx}
\usepackage{dcolumn}
\usepackage{bm}
\usepackage[usenames]{color}

\begin{document}

\title{Aligning spins in antiferromagnetic films using antiferromagnets}

\author{S. I. Csiszar}
 \affiliation{MSC, University of Groningen, Nijenborgh 4,
  9747 AG Groningen, The Netherlands}
\author{M. W. Haverkort}
 \affiliation{II. Physikalisches Institut, Universit\"{a}t zu K\"{o}ln,
  Z\"{u}lpicher Str. 77, 50937 K\"{o}ln, Germany}
\author{T. Burnus}
 \affiliation{II. Physikalisches Institut, Universit\"{a}t zu K\"{o}ln,
  Z\"{u}lpicher Str. 77, 50937 K\"{o}ln, Germany}
\author{Z. Hu}
 \affiliation{II. Physikalisches Institut, Universit\"{a}t zu K\"{o}ln,
  Z\"{u}lpicher Str. 77, 50937 K\"{o}ln, Germany}
\author{A. Tanaka}
 \affiliation{Department of Quantum Matter, ADSM, Hiroshima University,
  Higashi-Hiroshima 739-8530, Japan}
\author{H. H. Hsieh}
 \affiliation{Chung Cheng Institute of Technology,
  National Defense University, Taoyuan 335, Taiwan}
\author{H.-J. Lin}
 \affiliation{National Synchrotron Radiation Research Center,
  101 Hsin-Ann Road, Hsinchu 30077, Taiwan}
\author{C. T. Chen}
 \affiliation{National Synchrotron Radiation Research Center,
  101 Hsin-Ann Road, Hsinchu 30077, Taiwan}
\author{J. C. Cezar}
  \affiliation{European Synchrotron Radiation Facility, BP 220,
  38043, Grenoble, France}
\author{N. B. Brookes}
  \affiliation{European Synchrotron Radiation Facility, BP 220,
  38043, Grenoble, France}
\author{T. Hibma}
 \affiliation{MSC, University of Groningen, Nijenborgh 4,
  9747 AG Groningen, The Netherlands}
\author{L. H. Tjeng}
 \altaffiliation[Corresponding author:]{  tjeng@ph2.uni-koeln.de}
 \affiliation{II. Physikalisches Institut, Universit\"{a}t zu K\"{o}ln,
  Z\"{u}lpicher Str. 77, 50937 K\"{o}ln, Germany}

\date{\today}

\begin{abstract}
We have explored the possibility to orient spins in
antiferromagnetic thin films with low magnetocrystalline
anisotropy via the exchange coupling to adjacent antiferromagnetic
films with high magnetocrystalline anisotropy. We have used MnO
as a prototype for a system with negligible single-ion
anisotropy. We were able to control its spin direction very
effectively by growing it as a film on antiferromagnetic CoO
films with different predetermined spin orientations. This result
may pave the way for tailoring antiferromagnets with low
magnetocrystalline anistropy for applications in exchange bias
systems. Very detailed information concerning the exchange
coupling and strain effects was obtained from the Mn $L_{2,3}$
soft x-ray absorption spectroscopy.
\end{abstract}


\maketitle

The study of the exchange bias phenomena in multilayered magnetic
systems is a very active research field in magnetism, not the
least motivated by the high potential for applications in
information technology. Various combinations of antiferromagnetic
(AFM) and ferromagnetic (FM) thin film materials have been
fabricated and intensively
investigated.\cite{Nogues99,Berkowitz99} There seems to be an
agreement among the experimental and theoretical studies that the
largest exchange bias effects can be found in systems containing
AFMs with a high magnetocrystalline anisotropy, such as CoO. The
simple underlying idea is that the anisotropy helps to fix the
spin orientation in the AFM while switching the magnetization in
the FM.

\begin{figure}
\includegraphics[width=0.35\textwidth]{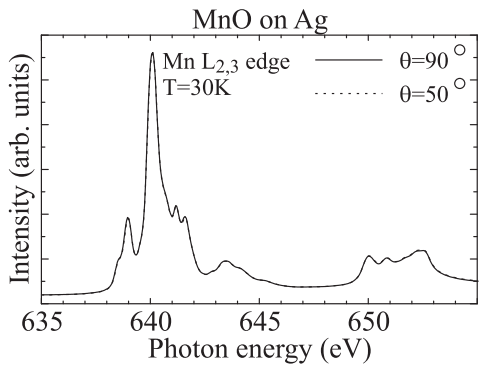}
     \caption{Mn $L_{2,3}$ XAS spectra of a 50 monolayer MnO film
     epitaxially grown in Ag(001). The spectra were taken with linearly
     polarized light, in which $\theta$ is the angle between
     the electric field vector $\vec{E}$ and the [001] surface normal.
     The sample was measured at 30 K.}
\end{figure}

Our objective is to explore the possibilities to control and to
pin the spin direction in AFM oxides having low
magnetocrystalline anisotropy, e.g. transition-metal oxides with
the $3d^{3}$, $3d^{5}$, or $3d^{8}$ ionic configurations. If
successful, this would help to extend the materials basis for the
AFMs used in exchange bias systems. One could then consider thin
films of not only NiO but also LaCrO$_{3}$, LaFeO$_{3}$,
$\gamma$-Fe$_{2}$O$_{3}$, and
R$_{3}$Fe$_{5}$O$_{12}$.\cite{Goodenoughbook} At first sight, our
chances may seem bleak since a recent study on ultra thin NiO
films reveals that the magnetic anisotropy results from a
detailed balance between the influence of strain and thickness on
the already very weak dipolar interactions in the
AFM.\cite{Altieri03,Finazzi03} Also a very recent detailed report
on MnO thin films grown on Ag(100) substrates\cite{Nagel07} did
not address the magnetic properties of the oxide film, suggesting
that it is difficult to detect magnetic order in those films.
Indeed, we ourselves have carried out at the ID08 beamline of the
ESRF in Grenoble polarization dependent x-ray absorption
spectroscopy (XAS) measurements at the Mn $L_{2,3}$ edges for a 50
monolayer (ML) thin film of MnO epitaxially grown on Ag(100) and,
as shown in Fig.1, we could not observe any linear dichroic
effect which otherwise could have unambiguously indicated the
presence of antiferromagnetic order like there is for NiO on
MgO.\cite{Alders95,Alders98} This is most probably the result of having several differently orientated anti-ferromagnetic domains in the sample, canceling the net linear dichroic effect. 

On the other hand, the few studies available in the literature on
combinations of AFM/AFM films revealed that the interlayer
exchange coupling can be very
strong.\cite{Ramos90,Carrico92,Wang92,Lederman93,Borchers93,Carey93,Abarra96}
We took these findings as starting point of our work. We have
used MnO as an ideal model for an antiferromagnetic system with
negligible single-ion anisotropy. For the antiferromagnetic
material with strong single-ion anisotropy we chose CoO. We note
that large magnetization loops shifts have been reported for MnO
nanocrystallites depositied by reactive ion beam on Co
metal,\cite{Lierop03} which we took as support for our hypothesis
that CoO/MnO interfaces may play an important role for the
magnetic coupling. We have grown MnO as a thin film epitaxially on
two different types of CoO single crystal films. In one CoO film
the spin direction is oriented perpendicular to the surface, and
in the other parallel to the surface.\cite{Csiszar05a} Using soft
x-ray absorption spectroscopy at the Mn $L_{2,3}$ edges, we will
show below that the spin direction of the MnO film strongly
depends on the type of CoO film the MnO is grown on, and that it
is dictated by the spin orientation of the CoO film and not by
the strain or dipolar interactions in the MnO film. Interlayer
exhange coupling is thus a very effective manner to control spin
directions and may be used for tailoring AFMs with low
magnetocrystalline anisotropy for exchange bias applications.

The actual MnO/CoO systems studied are
(14\AA)MnO/(10\AA)CoO/(100\AA)MnO/Ag(001) and
(22\AA)MnO/(90\AA)CoO/Ag(001). The two samples were grown on a
Ag(001) single crystal by molecular beam epitaxy (MBE),
evaporating elemental Mn and Co from alumina crucibles in a pure
oxygen atmosphere of $10^{-7}$ to $10^{-6}$ mbar. The base
pressure of the MBE system is in the low $10^{-10}$ mbar range.
The thickness and epitaxial quality of the films are monitored by
reflection high energy electron diffraction (RHEED) measurements.
Both the CoO and MnO films had [001]/[001] epitaxial
relationships with the underlying Ag(001) substrate. With the
lattice constants of bulk Ag, CoO, and MnO being 4.09\AA,
4.26\AA, and 4.444\AA, respectively, we find from x-ray
diffraction (XRD) and RHEED that the in-plane lattice constants
in each film are essentially given by the thickest layer which is
almost bulk like. Compared to the bulk, the 10\AA~CoO sandwiched
by MnO is about 4\% expanded in-plane ($a_{\parallel}\approx
4.424$\AA), while the 90\AA~CoO directly on Ag is slightly
compressed in-plane ($a_{\parallel}\approx4.235$\AA,
$a_{\perp}\approx4.285$\AA). The MnO is compressed in both
samples, but much more so for the one on the 90\AA~CoO film.
Details about the growth is given elsewhere.\cite{Csiszar05b} We
have recently shown that the spin direction is oriented
perpendicular to the surface in the CoO film under tensile
in-plane stress, and that it is parallel to the surface in the
film with the slightly compressive in-plane
stress.\cite{Csiszar05a}

\begin{figure}[h]
     \includegraphics[width=0.40\textwidth]{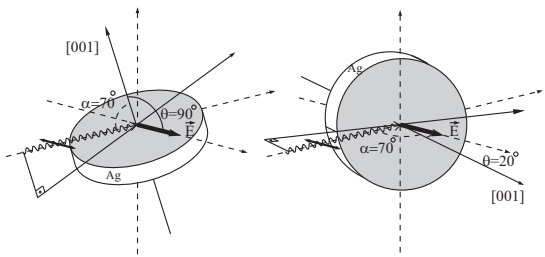}
     \caption{Experimental XAS geometry, with polarization
     of the light in the horizontal plane. $\theta$ is the angle
     between the electric field vector $\vec{E}$ and the [001] surface
     normal, and $\alpha$ the tilt between the Poynting vector and the
     surface normal.}
     \label{incidence}
\end{figure}

The XAS measurements were performed at the Dragon beamline of the
NSRRC in Taiwan using \textit{in-situ} MBE grown samples. The
spectra were recorded using the total electron yield method in a
chamber with a base pressure of 3$\times$10$^{-10}$ mbar. The
photon energy resolution at the Mn $L_{2,3}$ edges ($h\nu \approx
635-655$ eV) was set at 0.2 eV, and the degree of linear
polarization was $\approx 98 \%$. The sample was tilted with
respect to the incoming beam, so that the Poynting vector of the
light makes an angle of $\alpha=70^{\circ}$ with respect to the
[001] surface normal. To change the polarization, the sample was
rotated around the Poynting vector axis as depicted in Fig. 2, so
that $\theta$, the angle between the electric field vector
$\vec{E}$ and the [001] surface normal, can be varied between
$20^{\circ}$ and $90^{\circ}$. This measurement geometry allows
for an optical path of the incoming beam which is independent of
$\theta$, guaranteeing a reliable comparison of the spectral line
shapes as a function of $\theta$. A MnO single crystal is
measured \textit{simultaneously} in a separate chamber to obtain
a relative energy reference with an accuracy of better than 0.02
eV.

\begin{figure*}
     \includegraphics[width=0.80\textwidth]{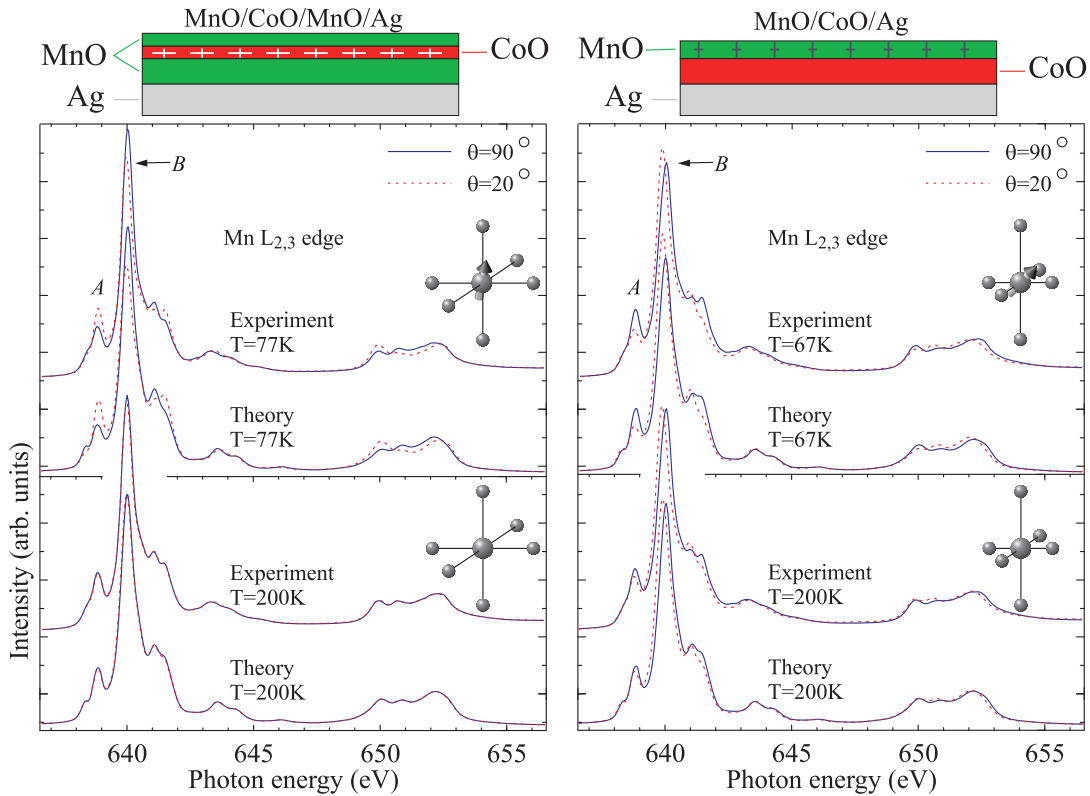}
     \caption{(color online) Experimental and calculated Mn $L_{2,3}$ XAS
     spectra of MnO in (left panel) (14\AA)MnO/(10\AA)CoO/(100\AA)MnO/Ag(001)
     and (right panel) (22\AA)MnO/(90\AA)CoO/Ag(001)
     for $\theta=20^{\circ}$ and $\theta=90^{\circ}$.
     Spectra below (T=77K left, T=67K right panel) and above (T=200K)
     the $T_{N}$ of the MnO film are shown in the top and bottom parts,
     respectively.}
\end{figure*}

Fig. 3 shows the polarization dependent Mn $L_{2,3}$ XAS spectra
of the MnO/CoO samples with CoO spin-orientation perpendicular
(left panel) and parallel (right panel) to the surface, taken at
temperatures far below (top part) and far above (bottom part) the
N\'{e}el temperature ($T_{N}$) of the MnO thin film, which is
about 130 K as we will show below. The spectra have been
corrected for electron yield saturation effects.\cite{Nakajima99}
The general line shape of the spectra shows the characteristic
features of bulk MnO,\cite{deGroot94} demonstrating the good
quality of our MnO films. Very striking in the spectra is the
clear polarization dependence, which is the strongest at low
temperatures. It is important to note that below $T_{N}$ the
dichroism, i.e. the polarization dependence, of the two samples
are opposite: for instance, the intensity of the peak $A$ at
$h\nu$ = 639 eV is higher for $\theta=20^{\circ}$ than for
$\theta=90^{\circ}$ in MnO/CoO where the spin orientation of the
CoO is out of plane, while it is smaller in the other sample.
Above $T_{N}$, the dichroism almost vanishes. Nevertheless, small
but clear and reproducible shifts in the spectra as a function of
polarization can be seen: peak $B$ at 640 eV has a shift of about
30 meV for the MnO sandwiching the 10\AA~CoO film, and 150 meV
for the MnO grown on top of the 90\AA~CoO film.

\begin{figure}
\includegraphics[width=0.35\textwidth]{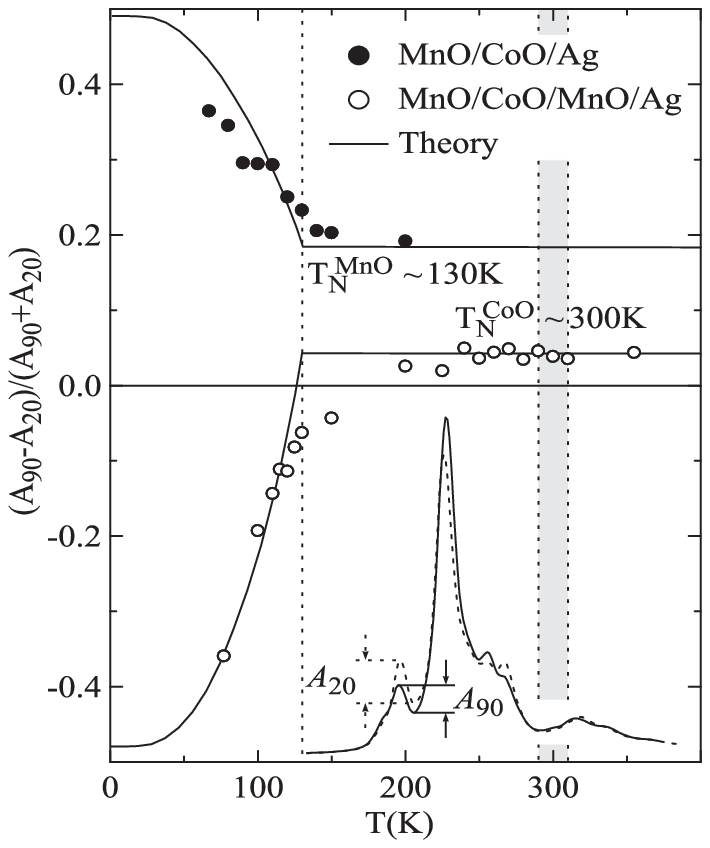}
     \caption{Temperature dependence of the polarization contrast
     in the Mn $L_{2,3}$ spectra, defined as the difference divided by the sum
     of the height of peak $A$ at $h\nu$ = 639 eV, taken with
     $\theta=20^{\circ}$ and $\theta=90^{\circ}$ polarizations. Filled and
     empty circles are the experimental data. The solid lines are the
     theoretical simulations. The shaded area represents the $T_{N}$
     of the CoO layers under the MnO film. }
\end{figure}

In order to resolve the origin of the dichroism in the spectra, we
have investigated the temperature dependence in more detail. The
measurements were done by taking the spectra starting from the
lowest temperature (T=67K or T=77K) towards higher temperatures
and after the highest temperature was reached (T=200K or T=400K)
we returned to low temperatures and acquired new spectra in
between the previously measured temperatures. Fig. 4 depicts the
polarization contrast of peak $A$, defined as the difference
divided by the sum of the peak height in the spectra taken with
the $\theta=20^{\circ}$ and $\theta=90^{\circ}$ polarizations. In
going from low to high temperatures, one can see a significant
temperature dependence for both samples (with opposite signs),
which flattens off at about 130K, identified as the $T_{N}$ of
these MnO thin films. We therefore infer that at low temperatures
the strong dichroic signal is caused by the presence of magnetic
ordering. Important is to note that the opposite sign in the
dichroism for the two samples implies that the orientation of the
magnetic moments is quite different. Here we recall that MnO
films grown on Ag \textit{without CoO} do not show any dichroism
down to the lowest temperatures, indicating the crucial role of
the CoO in orienting the magnetic moments of the MnO.

To understand the Mn $L_{2,3}$ spectra quantitatively, we
performed calculations for the atomic-like $2p^{6}3d^{5}
\rightarrow 2p^{5}3d^{6}$ transitions using a similar method as
described by Kuiper \textit{et al.}\cite{Kuiper93} and Alders
\textit{et al.}, \cite{Alders98} but now in a $D_{4h}$ point
group symmetry and including covalency. The method uses a MnO$_6$
cluster which includes the full atomic multiplet theory and the
local effects of the solid.\cite{deGroot94,Tanaka94} It accounts
for the intra-atomic $3d$-$3d$ and $2p$-$3d$ Coulomb and exchange
interactions, the atomic $2p$ and $3d$ spin-orbit couplings, the
O $2p$ - Mn $3d$ hybridization, local crystal field parameters
$10Dq$, $Ds$, and $Dt$, and a Brillouin-type temperature dependent
exchange field which acts on spins only and vanishes at $T_N$. As in earlier works, \cite{Alders98, Csiszar05a} an average has been made over all possible equivalent magnetic domains in the local $D_{4h}$ crystal-field point group, canceling out the effects of anisotropic magnetic linear dichroism. \cite{Arenholz06} 
The calculations were carried out using the XTLS 8.0
program.\cite{Tanaka94}

The results of the calculations are shown in Fig. 3. We used the
parameters already known for bulk MnO,\cite{Tanaka94,parameters}
and tuned only the parameters for $Ds$, $Dt$, and the direction of
the exchange field. For the MnO sandwiching the 10\AA~CoO we find
an excellent simulation of the experimental spectra for $Ds$ =
9.3 meV, $Dt$ = 2.6 meV, and an exchange field parallel to the
[112] direction. For the MnO overlaying the 90\AA~CoO we obtained
the best fit for $Ds$ = 48.6 meV, $Dt$ = 11.1 meV, and an
exchange field along the [211] direction. These two sets of
parameters reproduce the spectra at all temperatures extremely
well. This is also demonstrated in Fig. 4, showing the excellent
agreement between the calculated and measured temperature
dependence of the dichroism in peak $A$. Most important is
obviously the information concerning the spin direction that can
be extracted from these simulations. Earlier results have shown
that in-plane tensile strained CoO layers (c/a$<$1) have their spin
orientation parallel with the [001] direction while slightly
compressed CoO/Ag(001) layers (c/a$>$1) have spin orientation
perpendicular with the [001] direction. \cite{Csiszar05a} We thus find that
the magnetic moments in the MnO are oriented towards the surface
normal when it is grown on the CoO film which has the spin
direction perpendicular to the surface, and that it is lying
towards the surface when MnO is grown on the CoO film which has
the parallel alignment. In other words, it seems that bulk-like
MnO domains are stabilized that follow the CoO spin-direction as
much as possible.

\begin{figure}
\includegraphics[width=0.35\textwidth]{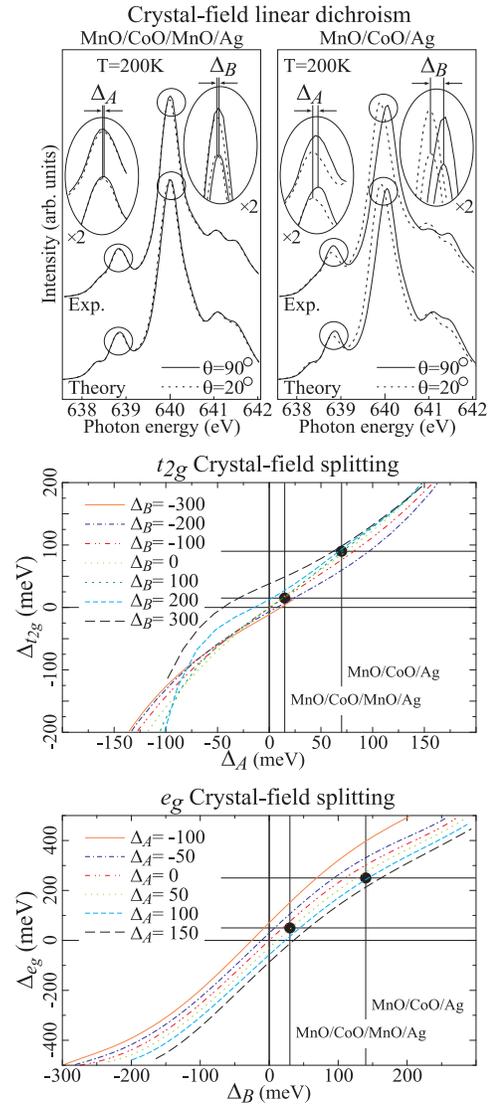}
 \caption{(color online) (top panel) Close-up of the experimental and
     calculated Mn $L_{3}$ XAS spectra at 200K, i.e. above
     $T_{N}$ of (left) (14\AA)MnO/(10\AA)CoO/(100\AA)MnO/Ag(001)
     and (right) (22\AA)MnO/(90\AA)CoO/Ag(001); $\Delta_A$ and
     $\Delta_B$ are the polarization dependent shifts in peaks $A$
     and $B$, respectively; (middle and bottom panels)
     Relationship between \{$\Delta_A$, $\Delta_B$\} and the
     tetragonal \{$\Delta_{e_g}$, $\Delta_{t_{2g}}$\} splittings.}
\end{figure}

In order to find out whether the spin direction in the MnO thin
films is determined by the exchange coupling with the CoO, or
whether it is given by strain and dipole interactions in the
films as found for NiO thin films on non-magnetic
substrates,\cite{Altieri03,Finazzi03} we now have to look more
closely into the tetragonal crystal fields in the MnO films. The
values for the tetragonal crystal field parameters $Ds$ and $Dt$,
which we have used to obtain the excellent simulations as plotted
in Fig. 3, can actually be extracted almost directly from the
high-temperature spectra, where the magnetic order has vanished
and does not contribute anymore to the polarization dependence.

The top panels of Fig. 5 show a close-up of the spectra taken at
200K, i.e. above $T_{N}$. One can now observe the small but clear
and reproducible shifts in the spectra as a function of
polarization: the shift in peak $A$ is denoted by $\Delta_A$ and
in peak $B$ by $\Delta_B$. In order to understand intuitively the
origin of these shifts, we will start to describe the energetics
of the high spin Mn$^{2+}$ ($3d^{5}$) ion in a one-electron-like
picture. In $O_{h}$ symmetry the atomic $3d$ levels are split
into 3 $t_{2g}$ and 2 $e_{g}$ orbitals, all containing a spin-up
electron. The $L_3$ edge of Mn$^{2+}$ should then consist of two
peaks: in exciting an electron from the $2p$ core level to the
$3d$, one can add an extra spin-down electron either to the lower
lying $t_{2g}$ or the higher lying $e_g$ shell, producing peaks
$A$ and $B$, respectively. In the presence of a tetragonal
distortion, both the $t_{2g}$ and $e_{g}$ levels will be split,
resulting in a polarization dependent shifts $\Delta_{A}$ and
$\Delta_{B}$, analogous as found for NiO.\cite{Haverkort04}

Due to the intra-atomic $2p$-$3d$ and $3d$-$3d$ electron
correlation effects, the relationship between the shifts in the
spectra and the crystal field splittings become non-linear. Using
the cluster model we are able to calculate this relationship for
a Mn$^{2+}$ $3d^{5}$ system and the results are plotted in Fig. 5.
Using this map, we find for the MnO sandwiching the 10\AA~CoO
that $\Delta_{t_{2g}}$ = 15 meV and $\Delta_{e_g}$ = 50 meV, and
for the MnO overlaying the 90\AA~CoO that $\Delta_{t_{2g}}$ = 90
meV and $\Delta_{e_g}$ = 250 meV
($Ds$=($\Delta_{e_g}$+$\Delta_{t_{2g}}$)/7,
$Dt$=(3$\Delta_{e_g}$-4$\Delta_{t_{2g}}$)/35). The crystal field
splittings for the second sample are much larger than for the
first sample. This is fully consistent with our structural data
which indicate that in the second sample MnO experiences a much
stronger compressive in-plane strain. It is important to
recognize that the crystal field splittings for the two samples
have the same sign, i.e. that both MnO films are compressed in
plane. This implies that strain together with the dipolar
interactions cannot explain the quite different spin-orientations
of the two MnO systems. We conclude that the magnetic anisotropy
mechanism present in, for instance, NiO thin films on
non-magnetic substrates,\cite{Altieri03,Finazzi03} is overruled
by the stronger interlayer exchange coupling
\cite{Ramos90,Carrico92,Wang92,Lederman93,Borchers93,Carey93,Abarra96}
between the CoO and MnO layers.

The $T_{N}$ for these thin MnO layers is found to be at about 130
K. It is surprising that the N\'{e}el temperature is not reduced
as compared to the bulk value of 121 K,\cite{Shull51} since
generally one would expect this to happen with decreasing
thickness as was observed for NiO on MgO.\cite{Alders98} The
origin for such a surprising behavior is not clear at this moment.
It is possible that the in-plane compressive stress causes an
increase in the Mn $3d$ - O $2p$ hybridization, which in turn
could produce an increase of the superexchange interaction
strength \cite{Goodenoughbook} and thus also of $T_{N}$. Another,
more exciting, possibility emerges from the recent experimental
and theoretical work on AFM/AFM multilayers such as
FeF$_{2}$/CoF$_{2}$ and
CoO/NiO.\cite{Ramos90,Carrico92,Wang92,Lederman93,Borchers93,Carey93,Abarra96}
Experiments have revealed that multilayers could even have a
single magnetic ordering transition temperature lying in between
the two $T_{N}$s of the constituent materials. The phenomenon has
been ascribed to the very strong interlayer exchange coupling.

In conclusion, we have shown that it is possible to control the
spin direction in MnO very effectively by growing them as thin
films on antiferromagnetic CoO films with different predetermined
spin orientations. Using detailed Mn $L_{2,3}$ soft x-ray
absorption spectroscopy, we are also able to show that it is not
strain but interlayer exchange coupling which plays a decisive
role herein. This result may pave the way for tailoring
antiferromagnets with low magnetocrystalline anistropy for
applications in exchange bias.

We acknowledge the NSRRC and ESRF staff for providing us with an
extremely stable beam. We would like to thank Lucie Hamdan and
Henk Bruinenberg for their skillful technical and organizational
assistance in preparing the experiment. The research in Cologne
is supported by the Deutsche Forschungsgemeinschaft through SFB
608.


\begin{thebibliography}{99}

\bibitem{Nogues99} See for review: J. Nogu\'{e}s and I. K. Schuller,
 J. Magn. Magn. Mater. \textbf{192}, 203 (1999).

\bibitem{Berkowitz99} See for review: A. E. Berkowitz and K. Takano,
 J. Magn. Magn. Mater. \textbf{200}, 552 (1999).

\bibitem{Goodenoughbook} J. B. Goodenough,
 \textit{Magnetism and the Chemical Bond} (Wiley, New York, 1963).

\bibitem{Altieri03} S.~Altieri, M.~Finazzi, H.~H.~Hsieh, H.-J.~Lin, C.~T.~Chen,
 T.~Hibma, S.~Valeri, and G.~A.~Sawatzky,
 Phys. Rev. Lett. \textbf{91}, 137201 (2003).

\bibitem{Finazzi03} M. Finazzi and S. Altieri,
 Phys. Rev. B \textbf{68}, 054420 (2003).

\bibitem{Nagel07} M.~Nagel, I.~Biswas, P.~Nagel, E.~Pellegrin, S.~Schuppler,
 H.~Peisert, and T.~Chass\'{e},
 Phys. Rev. B \textbf{75}, 195426 (2007).

\bibitem{Alders95} D.~Alders, J.~Vogel, C.~Levelut, S.~D.~Peacor, T.~Hibma,
 M.~Sacchi, L.~H.~Tjeng, C.~T.~Chen, G.~van~der~Laan, B.~T.~Thole, and G.~A.~Sawatzky,
 Europhys. Lett. {\bf 32}, 259 (1995).

\bibitem{Alders98} D.~Alders, L.~H.~Tjeng, F.~C.~Voogt, T.~Hibma,
 G.~A.~Sawatzky, J.~Vogel, M.~Sacchi, S.~Iacobucci, and C.~T.~Chen,
 Phys. Rev. B \textbf{57}, 11623 (1998).

\bibitem{Ramos90} C. A. Ramos, D. Lederman, A. R. King, and V. Jaccarino,
 Phys.Rev. Lett. \textbf{65}, 2913 (1990).

\bibitem{Carrico92} A. S. Carri\c{c}o and R. E. Camley,
 Phys. Rev. B \textbf{45}, 13117 (1992).

\bibitem{Wang92} R. W. Wang and D. L. Mills,
 Phys. Rev. B \textbf{46}, 11681 (1992).

\bibitem{Lederman93} D. Lederman, C. A. Ramos and V. Jaccarino, and J. L. Cardy,
 Phys. Rev. B, \textbf{48}, 8365 (1993).

\bibitem{Borchers93} J. A. Borchers, M. J. Carey, R. W. Erwin, C. F. Majkrzak, and A. E. Berkowitz,
 Phys. Rev. Lett. \textbf{70}, 1878 (1993).

\bibitem{Carey93} M. J. Carey, A. E. Berkowitz, J. A. Borchers, and R. W. Erwin,
 Phys. Rev. B \textbf{47}, 9952 (1993).

\bibitem{Abarra96} E. N. Abarra, K. Takano, F. Hellman, and A. E. Berkowitz,
 Phys. Rev. Lett. \textbf{77}, 3451 (1996).

\bibitem{Lierop03} J.~van~Lierop, M.~A.~Schofield, L.~H.~Lewis, R.~J.~Gambino,
 J. Magn. Magn. Mater. \textbf{264}, 146 (2003).

\bibitem{Csiszar05a} S.~I.~Csiszar, M.~W.~Haverkort, Z.~Hu, A.~Tanaka, H.~H.~Hsieh,
 H.-J.~Lin, C.~T.~Chen, T.~Hibma, and L.~H.~Tjeng,
 Phys. Rev. Lett. \textbf{95}, 187205 (2005).

\bibitem{Csiszar05b} S. I. Csiszar,
  thesis, University of Groningen (2005);
  http://irs.ub.rug.nl/ppn/287545644.

\bibitem{Nakajima99} R. Nakajima, J. St\"{o}hr and Y. U. Idzerda,
  Phys. Rev. B, \textbf{59}, 6421 (1999).

\bibitem{deGroot94} See review by F. M. F. de Groot,
  J. Electron Spectrosc. Relat. Phenom. {\bf 67}, 529 (1994).

\bibitem{Kuiper93} P. Kuiper, B. G. Searle, P. Rudolf, L. H. Tjeng, and C. T. Chen,
 Phys. Rev. Lett. \textbf{70}, 1549 (1993).

\bibitem{Arenholz06} E. Arenholz, G. van der Laan, R. V. Chopdekar, and Y. Suzuki, Phys. Rev. B \textbf{74}, 094407 (2006).

\bibitem{Tanaka94} A. Tanaka and T. Jo,
 J. Phys. Soc. Jpn. \textbf{63}, 2788 (1994).

\bibitem{parameters} Parameters for MnO$_{6}$ cluster [eV]:
$\Delta$=8.0, $U_{dd}$=5.5, $U_{\underline{c}d}$=7.2,
$V_{e_g}$=-2.1, $T_{pp}$=0.7, $10Dq$=0.5+70/12 $Dt$, $\zeta$=0.066,
$H_{ex}$=0.0135; Slater integrals 80\% of Hartree-Fock values;
MnO/CoO/MnO/Ag(100): $Ds$=0.0093, $Dt$=0.0026; MnO/CoO/Ag(100):
$Ds$=0.0486, $Dt$=0.0111. $H_{ex}$ from G. Pepy, J. Phys. Chem.
Solids \textbf{35}, 433 (1974).

\bibitem{Haverkort04} M.~W.~Haverkort, S.~I.~Csiszar, Z.~Hu, S.~Altieri,
 A.~Tanaka, H.~H.~Hsieh, H.-J.~Lin, C.~T.~Chen, T.~Hibma, and L.~H.~Tjeng,
 Phys. Rev. B \textbf{69}, 020408 (2004)

\bibitem{Shull51} C. G. Shull, W. A. Strauser, E. O. Wollan,
 Phys. Rev. \textbf{83}, 333 (1951).

\end{thebibliography}
\end{document}